\documentclass{ws-procs9x6}

\begin{document}

\title{SPIN FLIP LOSS IN MAGNETIC STORAGE OF ULTRACOLD NEUTRONS}

\author{A. STEYERL$^*$, C. KAUFMAN, G. M\"ULLER, S. S. MALIK, and A.M. DESAI}

\address{Department of Physics, University of Rhode Island,\\
Kingston, Rhode Island 02881, USA\\
$^*$asteyerl@mail.uri.edu\\
www.phys.uri.edu}

\begin{abstract}
We analyze the depolarization of ultracold neutrons confined in a magnetic field configuration similar to those used in existing or proposed magneto-gravitational storage experiments aiming at a precise measurement of the neutron lifetime. 
We use an approximate quantum mechanical analysis such as pioneered by Walstrom \emph{et al} [Nucl. Instrum. Methods Phys. Res. A \textbf{599}, 82 (2009)].  
Our analysis is not restricted to purely vertical modes of neutron motion.
The lateral motion is shown to cause the predominant depolarization loss in a magnetic storage trap.
\end{abstract}

%\keywords{...}

\bodymatter

%%%%%%%%%%%%%%%%%%%%%%%%%%%%%%%%%%%%%%%%%%%%%%%%%%%%%%
%
\section{Introduction}\label{sec:1} 
%
%%%%%%%%%%%%%%%%%%%%%%%%%%%%%%%%%%%%%%%%%%%%%%%%%%%%%%
The neutron lifetime $\tau_{\textrm{n}}$ is an important parameter in tests of the Standard Model of particle physics. 
It also affects the rate of helium production in the early universe and the energy production in the sun. 
The current Particle Data Group (PDG) average is $\tau_{\textrm{n}}$ =  880.1$\pm$ 1.1 s \cite{1}. 
However, the value of one experiment \cite{2}, which reported the lowest measurement uncertainty, namely $\sim$ 0.8 s, is located some 3.5 s below the bulk of other data in the PDG collection.\cite{3,4,5,6,7,8} 
The latter are consistently grouped around 882.0 s ($\pm$ 1.0 s) \cite{9}. 
As a possible way of advancing this field, storage of polarized ultracold neutrons (UCNs) in a magnetic trap has been pioneered by Paul {\it{et al.}} \cite{10} and is currently being pursued vigorously by several groups worldwide \cite{11,12,13,14,15}.
In magnetic traps there are no wall losses, the slow loss due to quasi-stable orbits is serious but believed to be manageable by avoiding regular orbits \cite{14}, and the potential loss due to depolarization, defined as spin flip relative to the local field direction, is argued to be negligible. 

Until recently UCN depolarization estimates \cite{16,17} have been based on Majorana's quasi-classical result of 1932 \cite{18} for a free polarized particle with magnetic moment moving with constant velocity vector through a non-uniform static magnetic field of specific form. 
For magnetic field parameters as currently used or proposed for UCN storage, $D$ would be of order $\exp(-10^{6})$, thus immeasurably small. Recently, Walstrom {\it{et al.}} \cite{14} pointed out that the values of $D$ for confined, rather than freely moving, neutrons are much larger. For a UCN moving along a vertical path in the storage system proposed by them, $D$ was estimated to be in the range $D \sim10^{-20}$ to $10^{-23}$. 
This is much larger than the Majorana value but still negligible in any actual or projected neutron lifetime experiment.

Using a simplified model of magnetic field distribution we extend that theory to include arbitrary UCN motion with both vertical and horizontal velocity components, confined to the vertical space between upper and lower turning points that depend only on the UCN energy for vertical motion. 
In our model (introduced in Sec.~\ref{sec:2}) the magnetic field magnitude $B$ is uniform within any horizontal plane, so there is no horizontal component of magnetic force. 
Therefore the neutron moves with constant velocity in the horizontal $z$- and $x$-directions. We show that $D$ could reach a level approaching the tolerance limit for a high-precision neutron lifetime measurement unless precautions are taken. 

Our model field is close to the ``bathtub configuration'' of Ref. \refcite{14} but the lateral confinement of UCNs, achieved there by double curvature of the magnetic mirror, is simulated differently. 
The magnetic mirror is horizontal and extends to infinity in both lateral dimensions. 
However, one could imagine the presence of ideal vertical mirrors reflecting the UCNs back and forth in the horizontal directions without any change in the analysis.
More specifically, we use an infinite ideal planar Halbach array\cite{19}, which is free of the field ripples present in actual realizations\cite{14}. 

The topic of UCN depolarization in magnetic storage or in mirror reflection in a magnetic field raises interesting questions of quantum interpretation. 
We postulate that the depolarization rate expected for a UCN magnetic storage experiment is determined by the current of UCNs in the ``wrong'' spin state.
Neutrons in this (high-field seeking) state exit the system at the lower and upper turning points, whereas neutrons  in the ``correct'' (high-field repelled) state are reflected and return to the storage space.
Exiting neutrons could be counted by detectors placed just outside the turning points.
In the Copenhagen interpretation, such a measurement (actual or hypothetical) resets the UCN wave function to a pure  state of high-field repelled neutrons. 
The spin state then evolves as described by the spin-dependent Schr\"odinger equation (or its semi-classical analog) until the next ``measurement'' takes place and the reset is repeated. 
A more comprehensive report of the present work can be found in Ref.~\refcite{22}, where we have also analyzed UCN depolarization in reflection from a nonmagnetic mirror placed into a nonuniform magnetic field.

%%%%%%%%%%%%%%%%%%%%%%%%%%%%%%%%%%%%%%%%%%%%%%%%%%%%%
%
\section{Magnetic field distribution}\label{sec:2} 
\begin{figure}[h]
  \begin{center}
 \includegraphics[width=90mm]{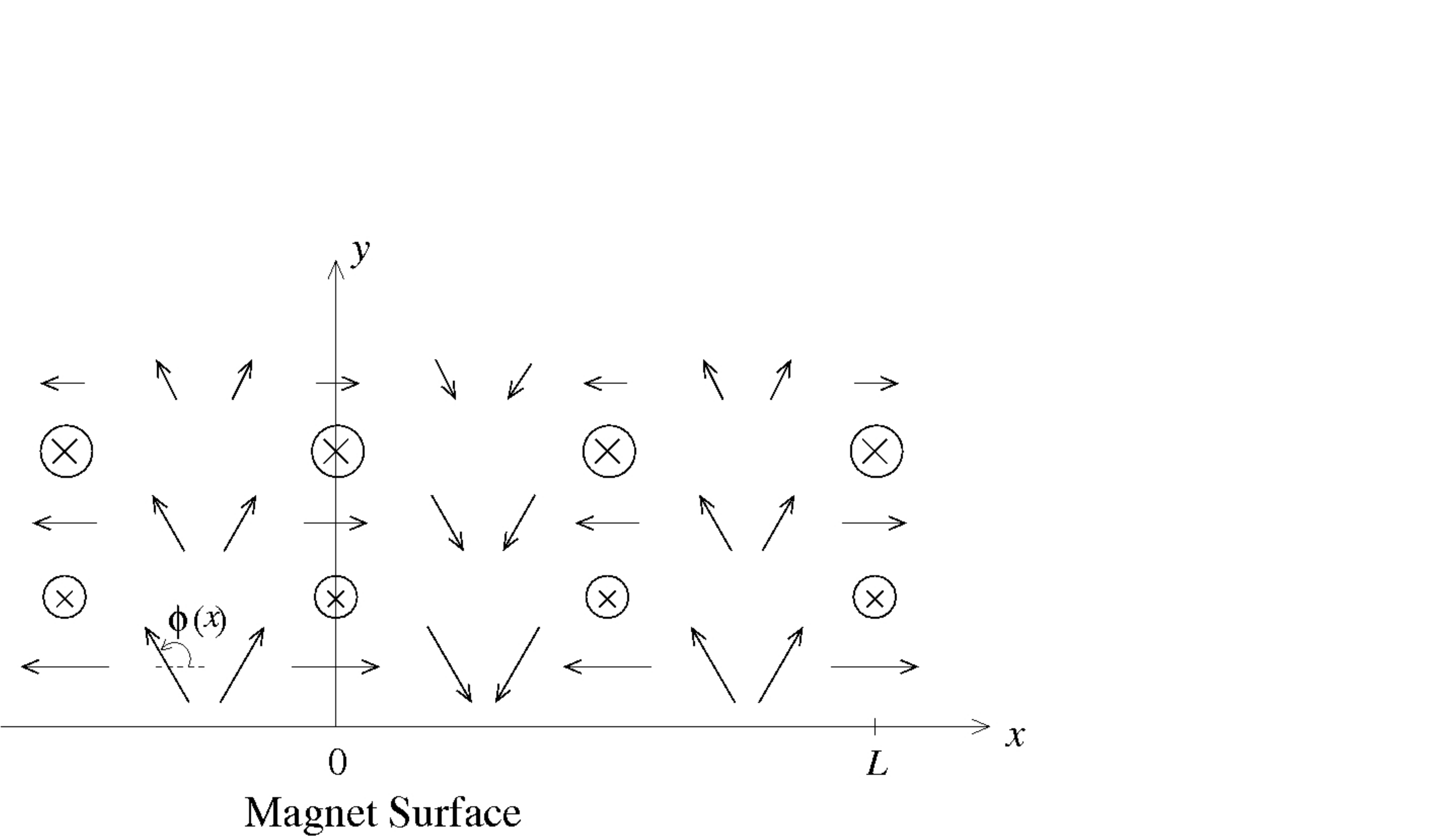}
\end{center}
\caption{The arrows show the Halbach magnetic field $\mathbf{B}_H$ as it rotates in the ($xy$)-plane.
Its magnitude $B_H$ decreases exponentially with height $y$ and is represented by the arrow length
using a log scale. 
The stabilization field $\mathbf{B}_1$ in the $z$-direction increases slowly with $y$ as symbolized by the crosses of variable size.}
  \label{fig:one}
\end{figure}

%
%%%%%%%%%%%%%%%%%%%%%%%%%%%%%%%%%%%%%%%%%%%%%%%%%%%%%
As illustrated in Fig.~\ref{fig:one}, an ideal Halbach array \cite{19} of permanent magnets of thickness $d$ covering the $(zx)$-plane generates a magnetic field
\begin{equation}\label{eq:1}
   {\bf{B}}_{H}(x,y) = B_0  {\mathrm{e}}^{-Ky} ({\mathbf{\hat x}} \cos Kx -{\mathbf{\hat y}} \sin Kx),
\end{equation}
where $B_0 = B_{\textrm{rem}} (1 - {\textrm{e}}^{-Kd})$ is determined by the remanent field $B_{\text{rem}}$.
We choose similar parameters as in Ref. \refcite{14}: $L=2\pi/K=5.2$cm, $d=2.54$cm and $B_0=0.82$T.
 
 %%%%%%%%%%%%%%%%%%%%%%%%%%%%%%%%%%%%%%%%%%%%%%%%%%
%%%%%%%%%%%%%%%%%%%%%%%%%%%%%%%%%%%%%%%%%%%%%%%%%%%

In the actual scheme \cite{14}, the uniform rotation is replaced by dividing the rotation period $L$ into four blocks, each of length $L/4$ and with the same magnetization $M$, but with an angle of 90$^\circ$ between the directions of ${\mathbf{M}}$ in adjacent blocks (schematically represented as  $...\leftarrow \downarrow\rightarrow\uparrow\leftarrow...)$.
Alternative designs are in the form of vertical or horizontal cylinders \cite{11,12,13,15}.

For the stabilization field we use  ${\mathbf{B}}_1 = \hat{\mathbf{z}} B_{10} \rho/(\rho - y)$ with $\rho = 1.5$ m. 
In Ref. \refcite{14} a value of 0.05  to 0.1 T was proposed for ${{B}}_{10}$. 
We consider similar field strengths down to the mT range.

%%%%%%%%%%%%%%%%%%%%%%%%%%%%%%%%%%%%%%%%%%%%%%%%%%%%%
%
 \section{Basic equations}\label{sec:3}   
%
%%%%%%%%%%%%%%%%%%%%%%%%%%%%%%%%%%%%%%%%%%%%%%%%%%%%%
The wave function for a UCN moving in the magneto-gravitational field of the trap is a linear superposition of the two eigenstates of the magnetic moment interaction Hamiltonian,
\begin{equation}\label{eq:2}
\mathcal{H}_m=-\mu_{\textrm{n}} {\mathbf{\sigma}}\cdot{\mathbf{B},}	
\end{equation}
where $\mu_{\textrm{n}}= -1.913  \mu_{\textrm{N}}$ is the neutron magnetic moment in terms of the
nuclear magneton $ \mu_{\textrm{N}}= 0.505\times10^{-26}$ J/T,  ${\mathbf{\sigma}}$ is the Pauli spin operator, and ${\mathbf{B}}$ is the local magnetic field. 
The two eigenstates $\chi^+$ and $\chi^-$ of $\mathcal{H}_m$ satisfy the eigenvalue equations,
\begin{equation}\label{eq:3}
\mathcal{H}_m \chi^{\pm}=\pm |\mu_{\textrm{n}}|B\chi^{\pm},
\end{equation}
with spin parallel and antiparallel to ${\mathbf{B}}$, respectively. 
These spin eigenfunctions are obtained by spin rotation from the $z$-axis to the direction of ${\mathbf{B}}$  through angles $\theta$  and $\phi$. 
The polar angle is  $\theta= \cos^{-1}(B_z/B) = \sin^{-1}(B_{xy}/B)$, where $B_z = B_1$ is due to the bias field ${\mathbf{B}}_1$ and $B_{xy} = B_H$ is the magnitude of the Halbach field ${\mathbf{B}}_H$. 
The azimuthal angle in the $(xy)$-plane is $\phi= \sin^{-1}(B_y/B_{xy}) = \tan^{-1}(B_y/B_x)$.

For the Halbach field configuration (\ref{eq:1}) we have $ \phi = -Kx$. 
Thus  $\phi$ depends only on $x$, while $\theta$ depends only on $y$.
Performing the spin rotation through angles $\theta$ and $\phi$ we obtain for the
spin basis vectors with quantization axis along ${\mathbf{B}}$,
\begin{equation}\label{eq:4}
\chi^- = \left( \begin{array}{c} e_-s  \\ -c  \\  \end{array} \right), \quad
\chi^+ = \left( \begin{array}{c} c  \\ e_+s \\  \end{array} \right),
\end{equation}
where $s = \sin(\theta/2)$, $c = \cos(\theta/2)$ and $e_{\pm} = \exp(\pm i\phi) = \exp(\mp iKx)$. 
We write the dependence of the wave function on position and spin in the form
\begin{equation}\label{eq:5}
 \chi= \alpha^{(3)}(x,y,z) \chi^+ +  \beta^{(3)}(x,y,z) \chi^-,
\end{equation}
where we have used the superscript ''(3)'' to indicate that $\alpha^{(3)}(x,y,z)$, $\beta^{(3)}(x,y,z)$ are functions of the three space coordinates.
By contrast, the functions $\alpha(y)$ and $\beta(y)$, introduced below, depend on $y$ only. 
$\chi$ satisfies the eigenvalue equation,
\begin{equation}\label{eq:6}
	E\chi =\left[ -\frac{\hbar^2}{2m}\nabla^2 + mgy + |\mu_{\textrm{n}}| {\mathbf{\sigma}}\cdot{\mathbf{B}}\right] \chi,
\end{equation}
for a neutron of mass $m$ with constant total energy $E$ moving in a uniform gravitational field of magnitude $g$ and a non-uniform magnetic field $\mathbf{B}$. 

As in Ref. \refcite{14} we use the WKB approximation \cite{23} and keep only terms that contain the derivatives of the field variables ($\theta$ and $\phi$) in lowest order.
Their variation is on the scale of centimeters whereas the waves in real space, $\alpha^{(3)}$ and $\beta^{(3)}$, vary on the micrometer scale, i.e. $\sim 10^4$ times faster. 
With the UCN initially in a pure (+) spin state, we obtain
\begin{align}\label{eq:7}
\nabla^2\chi &= (\alpha^{(3)}_{xx} + \alpha^{(3)}_{yy} + \alpha^{(3)}_{zz}) \chi^+ \nonumber \\
&+ [\beta^{(3)}_{xx}  +  \beta^{(3)}_{yy} + \beta^{(3)}_{zz} 
+ {\textrm{e}}^{-iKx} (-\theta_y \alpha^{(3)}_y + iK\alpha^{(3)}_x\sin\theta)] \chi^- ,
\end{align}
for scenarios where $|\beta^{(3)}|\ll |\alpha^{(3)}|$ holds. 
The expressions multiplying $\chi^+$ and $\chi^-$ can be simplified by noting that the $x$- and $z$-dependences of $\alpha^{(3)}$ have the plane-wave form $\mathrm{e}^{ik_xx}\mathrm{e}^{ik_zz}$ and $\beta^{(3)}$ is proportional to $\mathrm{e}^{-iKx}\mathrm{e}^{ik_xx}\mathrm{e}^{ik_zz}$. 
The wave numbers $k_x$ and $k_z$ are constant and $\mathrm{e}^{-iKx}$ represents a Bloch-wave modulation due to the periodicity of the Halbach field. In practice,  $k_x$ and $k_z$ are of order $\mu$m$^{-1}$, thus much larger than $K$ and $\theta_y$, both of which are of order cm$^{-1}$.
Thus we can factor Eq.~(\ref{eq:7}) in the form
\begin{align}\label{eq:8}
\nabla^2 \chi  &= \mathrm{e}^{ik_xx}\mathrm{e}^{ik_zz}\{[\alpha''  -  (k_x^2+k_z^2)\alpha]\chi^+ \nonumber \\
&+  \mathrm{e}^{-iKx}[\beta'' - (k_x^2+k_z^2)\beta - (\theta'\alpha'+Kk_x\alpha\sin\theta)]\chi^-\},
\end{align}
simplifying the notation. In Eq. (8) and henceforth, $\alpha(y)$ and $\beta(y)$ stand for the $y$-dependent parts of the wave function only, and differentiation with respect to $y$ is denoted by primes. 
We also drop the subscript $y$.
We thus write $\alpha^{(3)}(x,y,z) = \alpha(y)\mathrm{e}^{ik_xx}\mathrm{e}^{ik_zz}$ and $\beta^{(3)}(x,y,z) = \beta(y)\mathrm{e}^{-iKx}\mathrm{e}^{ik_xx}\mathrm{e}^{ik_zz}$. 

Substituting Eq.~(\ref{eq:8}) into the eigenvalue equation (\ref{eq:6}) gives \cite{14} two coupled equations, one for spinor $\chi^+$ and the other for $\chi^-$ :
 \begin{equation}\label{eq:9}
   E\alpha = -\frac{\hbar^2}{2m}\left[\alpha'' - (k_x^2+k_z^2)\alpha\right]+mgy\alpha+|\mu_{\textrm{n}}|B\alpha
\end{equation}
and
\begin{equation}\label{eq:10}
    E\beta=-\frac{\hbar^2}{2m}\left[\beta'' - (k_x^2+k_z^2)\beta-(\theta'\alpha'+Kk_x\alpha\sin\theta)\right]+mgy\beta-|\mu_n|B\beta.
\end{equation}
The WKB solution of (\ref{eq:9}) is\cite{14}
\begin{equation}\label{eq:11}
\alpha(y) = k_{+}^{-1/2}(y)\exp\Big(\pm i\Phi_+(y)\Big),
\end{equation}
where
\begin{equation}\label{eq:12}
\frac{\hbar^2k_{\pm}^2(y)}{2m}=E-\frac{\hbar^2}{2m}\left(k_x^2+k_z^2\right)+mg(y_0-y)\mp|\mu_{\textrm{n}}| B(y).
\end{equation}
Here $y_0$ is the greatest height a neutron of energy $E$ and given $k_x$ and $k_z$ would reach in the gravitational field if the magnetic field were switched off, and $\Phi_+(y) = \int_{y_s}^y k_+(u) du$ is the phase angle for the + spin state, accumulated between the start of vertical motion  and the position $y$. 
The initial height $y_s$ for motion upward is assumed to be that of the lower turning point, thus $y_{s^+} = y_l$,
and for motion downward the initial level is taken at the upper turning point, $y_{s^-} = y_u$. The additional + or $-$ sign in the argument of the exponential function in (\ref{eq:11}), in front of $\Phi_+$, refers to this direction of the motion; $+$ for upward and $-$ for downward, as in Ref. \refcite{14}.

The WKB wave function (\ref{eq:11}) is normalized to a constant particle flux $\hbar/m$ in $y$-direction. 
For the spin-flipped UCNs, the flux in the $y$-direction is the measure of the probability of depolarization, as shown below.
At the classical turning points, where $k_+ = 0$, the WKB form (\ref{eq:11}) diverges and has to be replaced by the Airy function\cite{14}.
Matching the Airy function to the WKB approximation is described in detail in Ref. \refcite{22}.

It follows from Eq.~(\ref{eq:10}) that the wave function $\beta(x,y)$ for the spin flipped component is determined by the inhomogeneous second-order differential equation,
\begin{equation}\label{eq:13}
\beta''(y) + k_-^2(y)\beta(y) = \theta'(y)\alpha'(y) + Kk_x \alpha(y) \sin\theta(y),
\end{equation}
and may be written \cite{14} in the WKB form
\begin{equation}\label{eq:14}
\beta(y) = k_{-}^{-1/2}(y)\exp\Big(\pm i\Phi_-(y)\Big) f(y),
\end{equation}
where the function $f(y)$ modulating the WKB wave represents the amplitude of spin flip. 
The phase accumulated since the start at a turning point, $\Phi_-(y) =\int_{y_s}^y  k_-(u) du$, always has a larger magnitude than the phase $\Phi_+(y)$ for $\alpha(y)$ since $k_-$ is greater than $k_+$ (except in zero magnetic field).

Thus the governing equation for $\beta(y)$ is the second-order differential equation
\begin{align}\label{eq:15}
\beta''(y) + k_-^2(y) \beta(y) =  [\pm ik_{+} \theta'(y) + Kk_x \sin\theta(y)] \alpha(y).
\end{align}
We have carried out the differentiation of $\alpha(y)$ using the WKB rule with the result $\alpha' = \pm ik_+\alpha(y)$, where the + sign applies to upward motion and the $-$ sign to downward motion. 
This replacement is valid except within a few $\mu$m of the turning points.

Our Eq.~(\ref{eq:15}) is consistent with Eq.~(28) of Ref. \refcite{14} except for the additional, $k_x$-dependent term on the right-hand side. 
It is present because we include motion with finite lateral momentum $\hbar k_x$. 
We will show that this new term dominates the UCN depolarization, since UCNs moving in $x$-direction are exposed to the strong field ripple due to the rotating Halbach field, in our model field as well as for the ``bathtub system'' \cite{14}.

%%%%%%%%%%%%%%%%%%%%%%%%%%%%%%%%%%%%%%%%%%%%%%%%%%%%%
%
\section{Depolarization in magnetic storage}\label{sec:4}  	
%
%%%%%%%%%%%%%%%%%%%%%%%%%%%%%%%%%%%%%%%%%%%%%%%%%%%%%
It has been shown in Ref. \refcite{22} that Eq.~(\ref{eq:15}) can be solved by straightforward integration.
For the downward motion we obtain
\begin{equation}\label{eq:16}
		\beta(y) = k_{-}^{-1/2}(y)P(y) \exp\Big(-i\Phi_+(y)\Big),
\end{equation}
where $ P(y) = [iU(y) + V(y)]/W(y)$, $U(y) = -\sqrt{k_+k_-} \theta'$, $V(y) = \sqrt{k_-/k_+} Kk_x \sin\theta$.
$W(y) = k_-^2(y) - k_+^2(y) = (4m/\hbar^2) |\mu_{\textrm{n}}| B(y)$ depends only on the magnitude $B(y)$ of the local magnetic field.

The phase $\Phi_+$ (with the index $+$) indicates that this wave for the ($-$) spin state propagates, not with wave number $k_-$, but with the same wave number $k_+$ as the (+) spin state, as  it should. 

Equation (\ref{eq:16}) represents a particular solution of (\ref{eq:15}) and we could add any solution $\beta_{h\pm}(y)$ of the homogeneous equation ${\beta}_{h}''(y) + k_-^2(y) \beta_h(y) = 0$. 
In the WKB framework, these solutions are $\beta_{h\pm}(y)=C_{\pm}k_{-}^{-1/2}(y)\exp\big(\pm i\Phi_{-}(y)\big)$ with arbitrary constants $C_{\pm}$. 
These functions represent a constant current in the upward (downward) direction for the + ($-$) sign. 
Thus the same current enters and exits the storage space, resulting in a zero contribution to the net flux out and, therefore to the depolarization. 

Reverting to solution (\ref{eq:16}), we identify the net depolarization over the path from upper turning point $y_u$ to $y_l$ as the current of spin-flipped UCNs at the endpoint $y_l$. 
This current represents the net flux out of the storage space since no flux enters at $y_u$.

At an arbitrary position $y$ along the way the current $j_-(y)$ is given by
\begin{align}\label{eq:17}
			j_-(y) &= \frac{\hbar}{m} \textrm{Re}\left[i {\beta}^*(y) \left(\frac{d\beta}{dy}\right)\right], \nonumber \\
 &= \frac{\hbar}{m} \left(\frac{k_+}{k_-}\right) |P|^2 = \frac{\hbar}{m} \frac{k_+^2 {\theta'}^2 + K^2k_x^2\sin^2\theta}{{(k_-^2 - k_+^2})^2}.
\end{align}

\begin{figure}[h]
  \begin{center}
 \includegraphics[width=100mm]{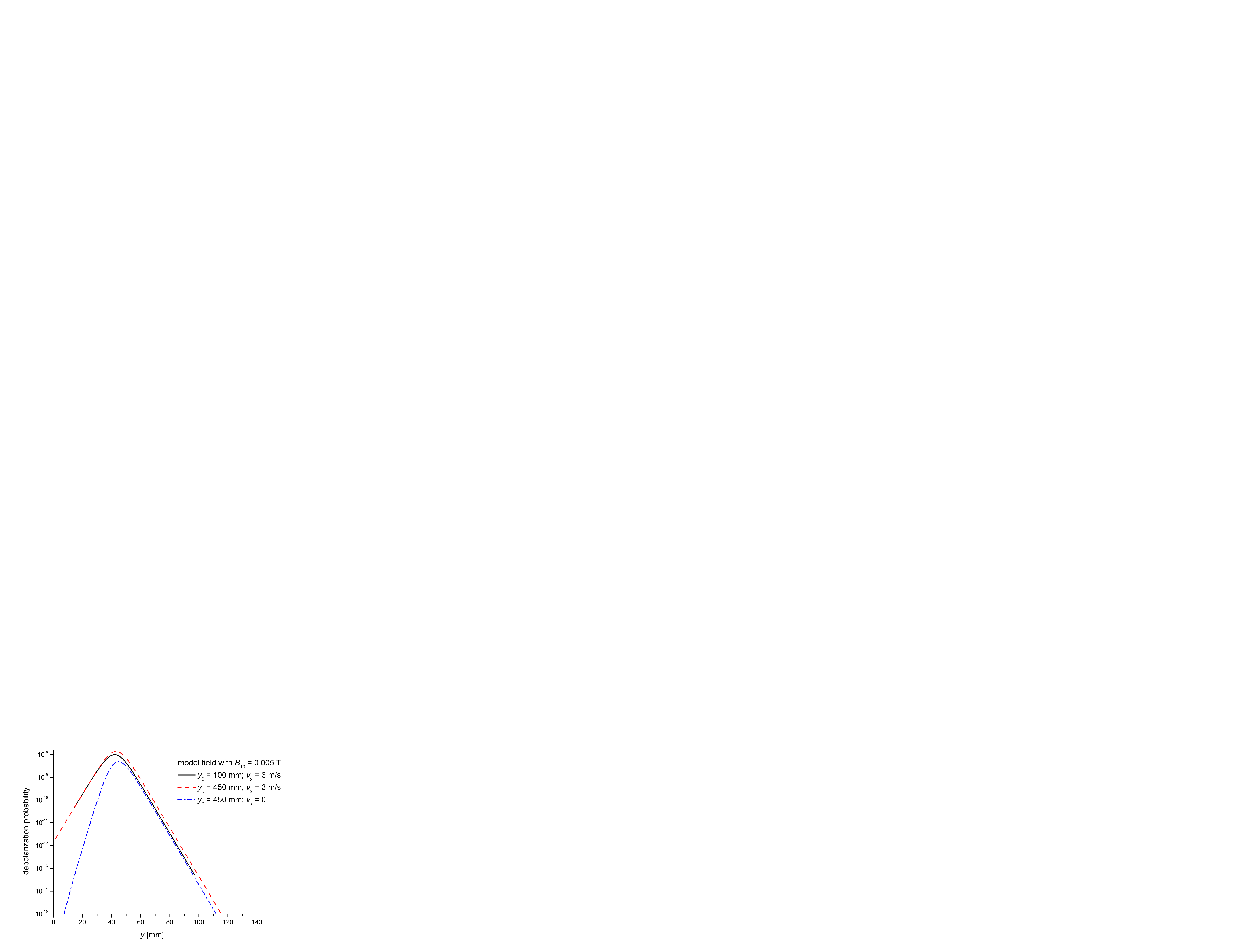}
\end{center}
\caption{Depolarization probability, given by Eq.~(\ref{eq:18}) multiplied by $m/\hbar$, as a function of neutron position for drop heights $y_0$ = 450 mm and 100 mm, stabilization field parameter $B_{10}$ = 0.005 T, and neutron velocity component $v_x$ = 3 m/s or zero. 
The sharp peak occurs in the region where the gradient of field angle $\theta$ is largest.}
  \label{fig:two}
\end{figure}

The depolarization probability $(m/\hbar) j_-(y)$ is plotted in Fig.~\ref{fig:two} for UCNs with energy for vertical motion determined by the ``drop heights'' $y_0 = 10$ cm and 45 cm, a bias magnetic field $B_{10}$ = 0.005 T and $v_x$ = 3 m/s. 
As in Ref. \refcite{14} we see a sharp peak at the $y$-position where $\theta'$ is large, and a decrease as the particle drops further down. The third curve in Fig.~\ref{fig:two} is for $y_0$ = 45 cm, $B_{10}$ = 0.005 T and $v_x$ = 0. 
The peak value and the decrease on the upper side are quite similar. Below the peak position the curve for $v_x$ = 0 decreases faster than for $v_x$ = 3 m/s.

 %%%%%%%%%%%%%%%%%%%%%%%%%%%%%%%%%%%%%%%%%%%%%%%%%%
%%%%%%%%%%%%%%%%%%%%%%%%%%%%%%%%%%%%%%%%%%%%%%%%%%%

The current leaving the storage space at $y = y_l$ is
\begin{equation}\label{eq:18}
j_l = \frac{\hbar}{m} \left(\frac{k_{+l}}{k_{-l}}\right) |P_l|^2 = \frac{\hbar}{m} \frac{k_{+l}^2 {{\theta}'_l}^2 + K^2k_x^2\sin^2\theta_l }{(k{_{-l}}^2 - k_{+l}^2)^2}= \frac{\hbar}{m} \frac{K^2k_x^2}{k_{-l}^4}\sin^2\theta_l ,
\end{equation}
where the index $l$ refers to the values at $y = y_l$ and the last expression uses the fact that $k_+$ vanishes at the turning points. 

The dependence of (\ref{eq:18}) on the field variables is established by noting that $\sin^{2}\theta =B_{H}^{2}/B^{2}$, $k_{-l}^{4}\sim B_{l}^{2}$ and $K^{2}k_{x}^{2}=(m/\hbar)^{2}\omega^{2}$, where $\omega = 2\pi v_{x}/L$ is the frequency of the Halbach field as seen by the moving UCN.

For upward motion from $y_l$ to $y_u$ we get the result for the current (18) with all indices $l$ replaced by $u$.
The quantities relevant for the spin-flipped current leaving the system at the upper turning point are determined by the field angle $\theta_u$ and by $k_{-u}$ at $y_u$.
The combined depolarization loss for one reflection on the magnetic field, i.e. for one complete round trip down and up thus becomes
\begin{equation}\label{eq:19}
\frac{m}{\hbar}  (j_l + j_u) = K^2k_x^2\left(\frac{\sin^2\theta_l}{k_{-l}^4} + \frac{\sin^2\theta_u}{k_{-u}^4}\right).
\end{equation}
In magnetic storage the UCNs have positive and negative velocities in any direction and, for a low-energy Maxwell spectrum, with uniform probability density in phase space. Thus we replace $k_x^2$ by its mean value,  $k_{x,max}^2/3$, for $-k_{x,\textrm{max}} < k_x < +k_{x,\mathrm{max}}$.

As a final step we establish the explicit connection between the mean loss current and the rate of depolarization, $\tau_{\textrm{dep}}^{-1}$, which should be negligible compared to the neutron $\beta$-decay rate in a neutron lifetime measurement.
For given neutron energy for vertical motion, i.e. fixed turning heights at $y_l$  and $y_u$, the depolarization rate (in s$^{-1}$) is determined by the loss current divided by the number of UCNs in the field-repelled spin state present in the trap,
\begin{equation}\label{eq:20}
N = 2 \int_{y_{l}}^{y_u} |\alpha(y)|^2 dy = 2 \int_{y_l}^{y_u} \frac{1}{k_+(y)} dy.
\end{equation}
We have used $|\alpha(y)|^2$ as the density and the factor 2 takes into account that both downward and upward moving UCNs are in the trap at the same time.

Since $k_+ = (m/\hbar) v_+$ and $dy = v_+ dt$, the expression in (\ref{eq:20}) equals $(\hbar/m)T$ where $T$ is the time required for one round trip down and up. 
Thus, the depolarization rate is
\begin{equation}\label{eq:21}
      \tau_{\textrm{dep}}^{-1} = \frac{\langle j_{l }+ j_{u}\rangle}{N} = \frac{m}{\hbar} \frac{\langle j_{l} + j_{u} \rangle}{T} = K^2 \left(\frac{k_{x,\textrm{max}}^2}{3}\right) \left(\frac{\sin^2\theta_ {l}}{k_{-l}^4} + \frac{\sin^2\theta_{u}}{k_{-u}^4}\right)\frac{1}{T}.
\end{equation}
This shows that the loss current (\ref{eq:19}) of spin-flipped UCNs equals the loss per round trip, i.e. for one bounce in the magnetic field.

For comparison with actual experiments we have to average (\ref{eq:21}) also over $v_y$.
As a measure of $v_y$ for a stored UCN we choose its value at the neutral plane $y = y^{(\textrm{n})}$, where the gravitational force is compensated by the magnetic force pushing upward, i.e. where $|\mu_{\textrm{n}}| (dB/dy) = -mg$. 
This is the plane where the UCNs with the lowest energy for vertical motion reside. In our field model, a UCN
with vertical velocity $v_y^{(n)} = 0$ in the neutral plane floats or moves along the plane at constant speed. 
In actual confinement fields as in Ref. \refcite{14} they would follow closed or open paths on the curved neutral surface.
For small values of $v_+^{(n)}$ the vertical motion is a classical harmonic oscillation with natural frequency $\omega_0 = \sqrt{dg_+/dy}$, where $g_+ = g + (|\mu_{\textrm{n}}|/m) (dB/dy)$ is the net downward acceleration.
This implies that for small oscillations about the neutral plane the time for a round trip becomes $T = 2\pi/\omega_0 = 2\pi(dg_+/dy)^{-1/2}$. 
For larger vertical velocities the oscillator potential is strongly anharmonic but the drop height $y_0$, used originally as a measure of energy for vertical motion, is unambiguously determined by $v_+^{(\textrm{n})}$. 
Therefore, if we plot the depolarization rate (\ref{eq:12}) versus $v_+^{(\textrm{n})}$, rather than $y_0$, the mean
height of this curve in the range from $v_+^{(n)} = 0$ to its maximum value for the stored UCN spectrum directly represents the average value of depolarization rate for a Maxwell spectrum. 

%%%%%%%%%%%%%%%%%%%%%%%%%%%%%%%%%%%%%%%%%%%%%%%%%%%%
\begin{figure}[h]
  \begin{center}
 \includegraphics[width=100mm]{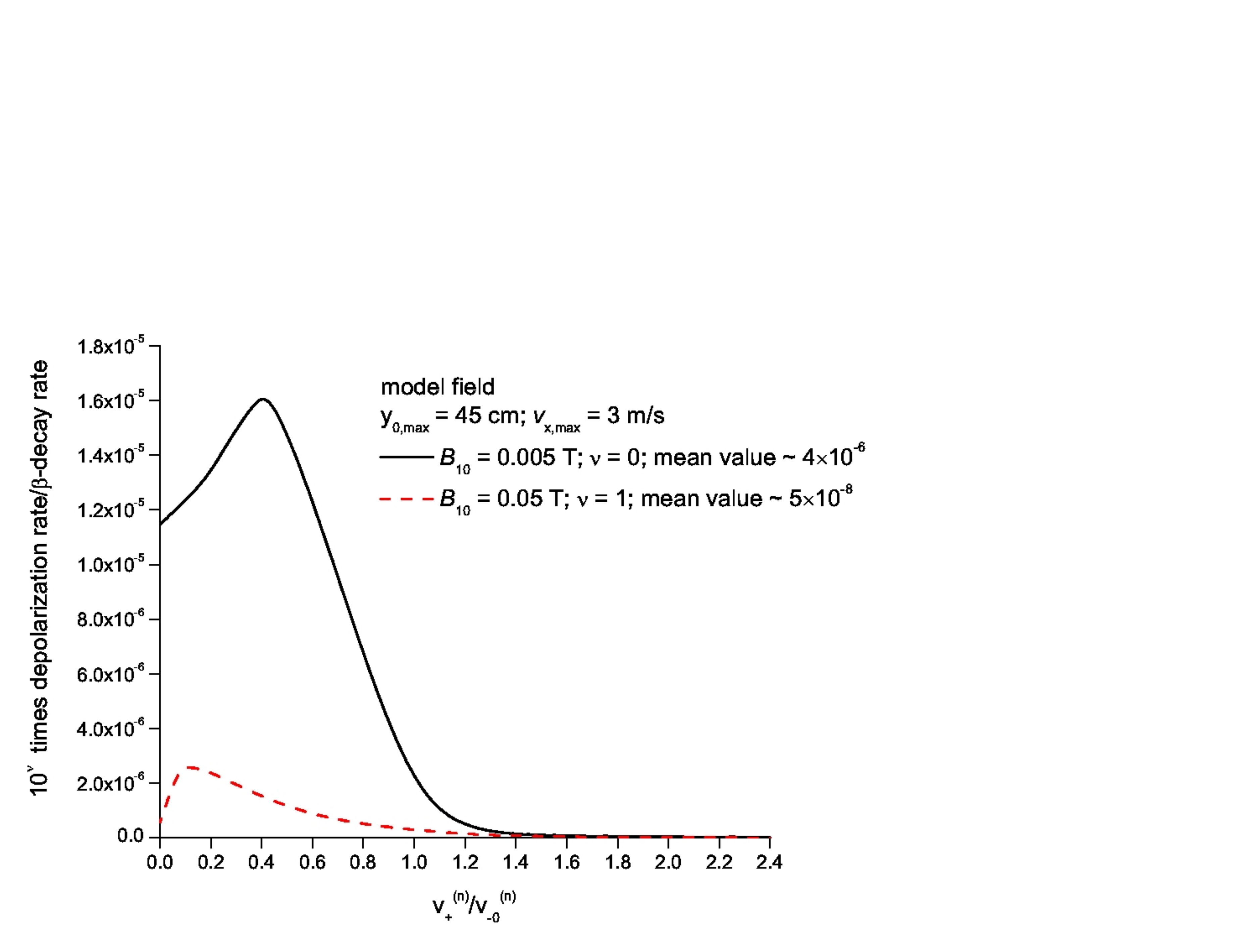}
\end{center}
\caption{Ratio between mean depolarization rate, given by Eq.~(\ref{eq:21}), and neutron $\beta$-decay rate, plotted as a function of vertical velocity component $v_+^{(\textrm n)}$ in the neutral plane (where the gravitational and magnetic forces are balanced). 
The curve for $B_{10}$ = 0.005 T is plotted to scale $(\nu= 0)$ and the curve for $B_{10}$ = 0.05 T is plotted with magnification factor $10^1$ $(\nu= 1)$.
Their difference by about two orders of magnitude shows the strong suppression of depolarization by a stabilization field of sufficient strength. 
For a Maxwell spectrum, the mean height of the curves over the range  of the abscissa, from 0 to 2.5 for $B_{10}=0.05$T and from 0 to 4.7 for $B_{10}=0.005$T,  directly determines the average over the full spectrum.}
  \label{fig:three}
\end{figure}
%%%%%%%%%%%%%%%%%%%%%%%%%%%%%%%%%%%%%%%%%%%%%%%%%%%%

Such a plot is presented in Fig.~\ref{fig:three}, where we have normalized $v_+^{(\textrm{n})}$ to  $v_{-0}^{(n)} = 2 \sqrt{|\mu_{\textrm{n}}|B^{(\textrm{n})}/m}$, the  $y$-velocity for the spin-flipped state on the neutral plane for  $v_+^{(\textrm{n})}=0$.
For $y_{0,\textrm{max}} = 45$ and $v_{x,\textrm{max}} = 3$ m/s the mean depolarization rate, normalized to the $\beta$-decay rate $1/\tau_{\textrm{n}}$, is $\tau_{\textrm{n}}\langle \tau_{\textrm{dep}}^{-1}\rangle = 4\times 10^{-6}$ for $B_{10}=0.005$ T and about two orders of magnitude less for $B_{10}=0.05$T.

The largest contribution to depolarization originates from UCNs with fairly low energy of vertical motion. They move through the field almost horizontally, with small vertical oscillations about the neutral plane. 
The result is plausible since these UCNs spend the largest fraction of time in the region where the field rotates rapidly in the moving reference frame.

%%%%%%%%%%%%%%%%%%%%%%%%%%%%%%%%%%%%%%%%%%%%%%%%%%%%%
%
\section{Conclusion}\label{sec:5}  	
%
%%%%%%%%%%%%%%%%%%%%%%%%%%%%%%%%%%%%%%%%%%%%%%%%%%%%%
We extend the analysis of Ref. \refcite{14} to include arbitrary UCN orbits with lateral velocity components.
As a main result of the extension we find that the lateral $x$-component of motion in the plane of the Halbach field makes the dominant contribution to depolarization while the depolarization due to the vertical motion is insignificant.
As a result, some previous estimates of depolarization probability may have been overoptimistic. 
For the parameters of \cite{14} (0.05-0.1 T for $B_{10}$) we estimate on the basis of Fig.~\ref{fig:three} that even a measurement of the neutron lifetime with precision $10^{-5}$ should be possible (disregarding other potential limitations) but the safety margin may be smaller than previously expected.

%%%%%%%%%%%%%%%%%%%%%%%%%%%%%%%%%%%%%%%%%%%%%%%%%%%%%
%
\section*{Acknowledgments}
%
%%%%%%%%%%%%%%%%%%%%%%%%%%%%%%%%%%%%%%%%%%%%%%%%%%%%%
We are grateful to R. Golub for useful comments and to V. Ezhov, C. Liu, and A. Young for having drawn our attention to depolarization in magnetic UCN confinement.

\bibliographystyle{ws-procs9x6}
\bibliography{ws-pro-sample}

\begin{thebibliography}{100}

\bibitem{1}
K.~Nakamura \textit{et al}.,
\newblock (Particle Data Group),
\newblock {\it J. Phys. G} \textbf{37}, 075021 (2010),
\newblock and 2012 partial update.

\bibitem{2}
A.~Serebrov \textit{et al}.,
\newblock {\it Phys. Lett. B}
\newblock\textbf{605}, 72 (2005);
\newblock {\it Phys. Rev. C}
\newblock\textbf{78}, 035505 (2008).

\bibitem{3}
W.~Mampe, P.~Ageron, C.~Bates, J.~M.~Pendlebury and A.~Steyerl,
\newblock {\it Phys. Rev. Lett.}
\newblock\textbf{63}, 593 (1989);
\newblock and update, Ref. [9].


\bibitem{4}
W.~Mampe, L.~N.~Bondarenko, V.~I.~Morozov, Yu.~N.~Panin and A.~I.~Fomin,
\newblock {\it JETP Lett.}
\newblock\textbf{57}, 82 (1993).


\bibitem{5}
J.~Byrne \textit{et al}.,
\newblock {\it Eur. Phys. Lett.}
\newblock\textbf{33}, 187 (1996).


\bibitem{6}
S.~S.~Arzumanov \textit{et al}.,
\newblock {\it Nucl. Instrum. Methods Phys. Res. A}
\newblock\textbf{440}, 511 (2000);
\newblock and update, {\it JETP Lett.} \textbf{95}, 224 (2012).


\bibitem{7}
J.~S.~Nico \textit{et al}.,
\newblock {\it Phys. Rev. C}
\newblock\textbf{71}, 055502 (2005).


\bibitem{8}
A.~Pichlmaier, V.~Varlamov, K.~Schreckenbach and P.~Geltenbort,
\newblock {\it Phys. Lett. B}
\newblock\textbf{693}, 221 (2010).


\bibitem{9}
A.~Steyerl, J.~M.~Pendlebury, C.~Kaufman, S.~S.~Malik, A.~M.~Desai,
\newblock {\it Phys. Rev. C}
\newblock \textbf{85}, 065503 (2012).



\bibitem{10}
W.~Paul \textit{et al}.,
\newblock {\it Z. Physik C}
\newblock\textbf{45}, 25 (1989).


\bibitem{11}
V.~F.~Ezhov \textit{et al}.,
\newblock {\it J. Res. Mat. Inst. Standards and Technology}
\newblock\textbf{110}, 1 (2005); V. F. Ezhov \textit{et al}.,
\newblock {\it Nucl. Instrum. Methods Phys. Res. A}
\newblock \textbf{611}, 167 (2009)


\bibitem{12}
K.~Leung, O.~Zimmer,
\newblock {\it Nucl. Instrum. Methods Phys. Res. A}
\newblock\textbf{611}, 181 (2009).


\bibitem{13}
P.~Huffman \textit{et al}.,
\newblock {\it Nature}
\newblock\textbf{403}, 62 (2000).



\bibitem{14}
P.~L.~Walstrom \textit{et al}.,
\newblock {\it Nucl. Instrum. Methods Phys. Res. A}
\newblock \textbf{599}, 82 (2009).


\bibitem{15}
S.~Materne \textit{et al}.,
\newblock {\it Nucl. Instrum. Methods Phys. Res. A}
\newblock\textbf{611}, 176 (2009).




\bibitem{16}
V.~V.~Vladimirsky,
\newblock {\it JETP}
\newblock\textbf{12}, 740 (1961).
	

\bibitem{17}
Yu.~N.~Pokotilovski,
\newblock {\it JETP Lett.}
\newblock\textbf{76}, 131 (2002); \textit{Erratum},
\newblock {\it JETP Lett.}
\newblock\textbf{78}, 422 (2003).


\bibitem{18}
E.~Majorana,
\newblock {\it Il Nuovo Cimento}
\newblock\textbf{9}, 43 (1932).



\bibitem{19}
J.~C.~Mallinson,
\newblock {\it IEEE Transactions on Magnetics}
\newblock\textbf{9}, 1 (1973).



\bibitem{20}
R.~W.~Pattie \textit{et al}.,
\newblock {\it Phys. Rev. Lett.}
\newblock\textbf{102}, 012301 (2009).


\bibitem{21}
J.~Liu  \textit{et al}.,
\newblock {\it Phys. Rev. Lett.}
\newblock\textbf{105}, 181803 (2010).


\bibitem{22}
A.~Steyerl, C.~Kaufman, G.~M\"uller, S.~S.~Malik, and A.~M.~Desai, 
\newblock {\it Phys. Rev. C}
\newblock \textbf{86}, 065501 (2012).

\bibitem{23}
P.~M.~Morse and H.~Feshbach,
\newblock \textit{Methods of Theoretical Physics}
\newblock (McGraw-Hill, New York, 1953), Chap. 9.3.

\end{thebibliography}

\end{document}